\documentclass[aps,prb,reprint,superscriptaddress]{revtex4-2}

\pdfoutput=1
\usepackage[utf8]{inputenc}
\usepackage{amsmath,amssymb,amsfonts,graphicx,verbatim,hyperref, mathtools, bbold, float}
\usepackage{physics}
\usepackage{soul, color}
\usepackage{algorithm, algorithmic}
\hypersetup{
	colorlinks=true,
	linkcolor=red,
	citecolor=blue,
	filecolor=blue,
}

\soulregister\ref{7}  
\soulregister\cite{7} 

\renewcommand*{\eqref}[1]{Eq.~(\ref{#1})}

\newcommand*{\figref}[1]{Fig.~\ref{#1}}

\newcommand*{\secref}[1]{Sec.~\ref{#1}}

\newcommand*{\appref}[1]{Appendix~\ref{#1}}

\begin{document}


\title{Temperature Dependent Energy Diffusion in Chaotic Spin Chains}
\author{Cristian \surname{Zanoci}}
\email{czanoci@mit.edu}
\affiliation{Department of Physics, Massachusetts Institute of Technology, 77 Massachusetts Avenue, Cambridge, MA 02139, USA}
\author{Brian \surname{Swingle}}
\email{bswingle@umd.edu}
\affiliation{Condensed Matter Theory Center and Joint Quantum Institute, Department of Physics, University of Maryland, College Park, Maryland 20742, USA}
\affiliation{Department of Physics, Brandeis University, Waltham, Massachusetts 02453, USA}

\begin{abstract}

We study the temperature dependence of energy diffusion in two chaotic gapped quantum spin chains, a tilted-field Ising model and an XZ model, using an open system approach. We introduce an energy imbalance by coupling the chain to thermal baths at its boundary and study the non-equilibrium steady states of the resulting Lindblad dynamics using a matrix product operator ansatz for the density matrix. We define an effective local temperature profile by comparing local reduced density matrices in the steady state to those of a uniform thermal state. We then measure the energy current for a variety of driving temperatures and extract the temperature dependence of the energy diffusion constant. For the Ising model, we are able to study temperatures well below the energy gap and find a regime of dilute excitations where rare three-body collisions control energy diffusion. A kinetic model correctly predicts the observed exponential increase of the energy diffusion constant at low temperatures. For the XZ model, we are only able to access intermediate to high temperatures relative to the energy gap and we show that the data is well described by an expansion around the infinite temperature limit. We also discuss the limitations of the particular driving scheme and suggest that lower temperatures can be accessed using larger baths.

\end{abstract}

\maketitle

\section{Introduction}
\label{sec:intro}

Understanding the physics of quantum systems out of equilibrium is a central challenge in many areas of physics, with applications ranging from solid state systems to the quark-gluon plasma. One key question is how to describe the macroscopic motion of energy and other conserved quantities starting from the microscopic physics. The study of such transport phenomena can reveal fundamental properties of quantum matter, such as superconductivity or conductance quantization, and gives insight into the underlying dynamical processes at work.

On general grounds, one expects a hydrodynamic description of the motion of conserved quantities at long distances and times in chaotic systems. However, computing the parameters of the hydrodynamic theory is a challenging problem that requires controlling the microscopic physics. Even for one-dimensional systems with local interactions, where the corresponding equilibrium problem is under much better control, the calculation of transport properties remains a difficult task~\cite{bertini2020finite}. While increases in computational power and numerical method developments enable progress~\cite{vidal2004efficient,verstraete2004matrix,zwolak2004,cui2015variational,mascarenhas2015matrix,gobert2005,langer2009real,weimer2019simulation}, there are also conceptual issues at play. The most fundamental of these is the need for a proper matching between the emergent hydrodynamic description at long scales and the exact quantum description at short scales~\cite{bertini2016,castro2016}. Simply put, hydrodynamics emerges at large size, but this is exactly where the brute force simulation of a quantum many-body system becomes impossible.

In the context of one-dimensional systems in equilibrium, matrix product state methods (and tensor network methods more generally)~\cite{schollwock2011density,PAECKEL2019167998} have proven extremely powerful for describing large-scale systems in equilibrium. However, for the kinds of chaotic systems expected to exhibit conventional diffusive transport, entanglement growth typically impedes the long-time simulation of such systems out of equilibrium. There are various ways to deal with this impasse. One is to design modified dynamical laws that deviate from microscopic unitary evolution, but which hopefully retain the physics of interest while being more amenable to tensor network methods~\cite{Haegeman2011,leviatan2017quantum,white2018quantum,rakovszky2020dissipation}. Another approach is to modify the physical setup to reduce entanglement while accessing the same observables. Specifically, the same hydrodynamic coefficients that govern transport in an isolated system should also control the flow when the system is coupled to a drive at its boundaries. Because the explicit coupling of a system to a bath is expected to reduce the entanglement in the system, one might therefore be able to access transport physics in a low-entanglement simulation using an open system approach~\cite{prosen2009matrix,znidaric2010,vznidarivc2011transport,vznidarivc2011spin,znidaric2012,znidaric2013a,znidaric2013b,mendoza2013heat, mendoza2015, mendoza2019asymmetry}. For this second approach, while the entanglement plausibly remains low during time evolution, new challenges arise including the key question of how to drive the system to the desired non-equilibrium steady state (NESS) and how to characterize the steady state. 

Alongside these theoretical developments, groundbreaking advances have also been made in experimental realizations of low-dimensional systems. In particular, experiments with ultracold quantum gases provide an excellent setup for investigating the transport properties of these systems, due to their high degree of controllability~\cite{Bloch2008,bloch2012quantum}. Recently, different transport regimes have been observed in ultracold atom experiments~\cite{brown2019bad,nichols2019spin,guardado2020subdiffusion,Salomon2019,vijayan186,jepsen2020spin}. In addition, non-equilibrium setups where a mesoscopic channel is driven by a temperature or chemical potential imbalance at the boundaries have been studied with cold gases~\cite{brantut2012conduction,brantut2013thermoelectric,Stadler2012,brantut2013thermoelectric,lebrat2018,Krinner2017}. The latter are more in line with the setup considered in this work. 

In this paper, we study energy transport in one-dimensional systems driven out of equilibrium by a temperature imbalance at their two ends. We consider two non-integrable spin-1/2 Hamiltonians which are known to exhibit diffusive energy transport at high temperature, a tilted-field Ising model and an XZ model in a transverse field~\cite{prosen2009matrix,ye2019emergent}. We expect these models to be reasonably representative of generic non-integrable gapped interacting systems in one dimension. Our main focus is the temperature dependence of the energy diffusion constant, and we specifically aim at going to lower temperatures than previously accessed using tensor network methods. A key component of our analysis is the development of an accurate thermometer which assigns a local temperature profile to the system based on comparing local density matrices in the steady state with the corresponding density matrices in thermal equilibrium. This method has the virtue that it is guaranteed to give the correct temperature in the limit of vanishing drive; it is also simpler to implement than previous approaches that explicitly parameterized the local density matrix~\cite{mendoza2015}.

Once we obtain a reliable estimate for the local temperature profile, we are able to map out the temperature dependence of the energy diffusion constant in both models down to a minimum accessible temperature. For the Ising model, this minimum temperature is well below the gap and we find that the diffusion constant depends exponentially on the inverse temperature at low temperature. Moreover, the temperature dependence is consistent with a kinetic model in which three-body scattering is responsible for energy diffusion. This is because two-body collisions are unable to relax an energy current in 1D due to the constraints of momentum and energy conservation. This is related to the corresponding theoretical results for spin transport, where two-body collisions can relax the spin current~\cite{damle1998spin,damle2005universal}. For the XZ model, we find that the smallest effective temperature is still above the gap, so we probe the high temperature regime of transport in that model. We also compare our results to previous work using a different tensor network method based on a modification of the microscopic unitary dynamics and find excellent agreement~\cite{ye2019emergent}. For both models, the diffusion constant approaches a fixed value at high temperature. To the best of our knowledge, this work represents the first attempt to characterize the temperature dependence of diffusion constants (especially in the low-temperature regime) in the context of open-system dynamics. 

The ultimate limiting factor in our study is the minimum accessible temperature, which is simply the temperature below which the open system consistently failed to reach a sensible steady state on the timescales we considered. The physically meaningful minimum temperature cannot be simply read off from the bath parameters, since the target temperature of the bath is not equal to the effective temperature induced in the system. Moreover, as we show, the smallest effective temperature we achieve can be either above or below other important scales, such as the energy gap. What exactly determines this limit on the temperature remains poorly understood. Here we restricted our attention to two-site baths, but we expect that lower temperatures can be accessed using a larger bath~\cite{zanoci2016entanglement}.

The outline of this paper is as follows. In \secref{sec:models} we define the models of interest. In \secref{sec:methods} we outline our methods and describe in detail our approach to thermometry. In \secref{sec:results} we present our results for both models, while focusing specifically on the low temperature regime. Finally, we give a short discussion and outlook in \secref{sec:discussion}.
\section{Models}
\label{sec:models}

We study the transport properties of two different one-dimensional spin-1/2 lattice models. For future convenience, we decompose the Hamiltonians in terms of bond operators $H=\sum_{i=1}^{L-1} H_{i, i+1}$ acting on sites $(i, i+1)$. First, we consider the Ising model in the presence of both transverse and longitudinal magnetic fields

\begin{equation}
    H_1 = \sum_{i=1}^{L-1} \Big(J_z \sigma_i^z\sigma_{i+1}^z + \frac{h_x}{2}(\sigma_i^x+\sigma_{i+1}^x)+\frac{h_z}{2}(\sigma_i^z+\sigma_{i+1}^z)\Big),
    \label{eq:model_1}
\end{equation}
where $\sigma_i^{x, z}$ are Pauli matrices at site $i$ and $L$ is the total length of the chain. This model is also known as the tilted-field Ising model. We choose the Hamiltonian parameters to be $(J_z, h_x, h_z) = (-2, 3.375, 2)$, for which transport at high temperatures has been previously studied using open-system dynamics~\cite{prosen2009matrix}. 

Our second model is an XZ spin chain in a transverse magnetic field

\begin{equation}
    H_2 = \sum_{i=1}^{L-1} \Big(J_x \sigma_i^x\sigma_{i+1}^x + J_z \sigma_i^z\sigma_{i+1}^z + \frac{h_x}{2}(\sigma_i^x+\sigma_{i+1}^x)\Big),
    \label{eq:model_2}
\end{equation}
which is just a transverse-field Ising model with an additional $\sigma_i^x\sigma_{i+1}^x$ coupling. Following Ref.~\cite{ye2019emergent}, we set $(J_x, J_z, h_x) = (1, 0.75, 0.21)$.  

We are interested in the transport of conserved quantities and both of these Hamiltonians have the energy as their only conserved quantity, since there is no $U(1)$ symmetry. The formulas for the associated local energy current $j_i$ can be derived from combining the continuity equation equation at site $i$ with Heisenberg's equation of motion~\cite{zotos1997transport} 

\begin{equation}
    j_i = i[H_{i-1, i}, H_{i, i+1}].
    \label{eq:current}
\end{equation}
In the long-time limit, when the system reaches its NESS, the current becomes uniform throughout the chain $j\equiv\langle j_i\rangle$. For the specific models defined above, the energy currents take on the form 

\begin{equation}
    j_1 = - h_xJ_z\langle(\sigma_{i-1}^z\sigma_i^y - \sigma_i^y\sigma_{i+1}^z)\rangle,
    \label{eq:current_1}
\end{equation}
\begin{equation}
\begin{split}
    j_2 &= 2J_xJ_z\langle(\sigma_{i-1}^x\sigma_i^y\sigma_{i+1}^z - \sigma_{i-1}^z\sigma_i^y\sigma_{i+1}^x)\rangle \\
    &- h_xJ_z\langle(\sigma_{i-1}^z\sigma_i^y - \sigma_i^y\sigma_{i+1}^z)\rangle,
    \label{eq:current_2}
\end{split}
\end{equation}
where the subscripts now refer to the two models, instead of a particular site. 

For the selected regime of parameters, both models are known to be non-integrable, quantum chaotic, and exhibit diffusive energy transport~\cite{prosen2009matrix,ye2019emergent}. Such systems obey Fourier's law, which states that the current is proportional to the gradient of the energy density $j=-D\nabla E = -D\Delta E/L$, where the proportionality factor is the diffusion constant and $\Delta E$ denotes the energy difference between the two ends of the chain. More generally, if transport is not diffusive, the current will scale with system size as $j\sim 1/L^\gamma$. Depending on the exponent $\gamma$, one can have (i) ballistic transport ($\gamma = 0$), (ii) superdiffusive transport ($0<\gamma<1$), (iii) diffusive transport ($\gamma = 1$), or (iv) subdiffusive transport ($\gamma > 1$)~\cite{bertini2020finite}. We will explicitly verify in our numerical simulations that the transport is indeed diffusive at all temperatures (see \secref{sec:thermal_results}). 

As we will show in \secref{sec:temp_dep_results}, the scaling of the diffusion constant at low temperatures depends on the energy gap $\Delta$, which sets the effective energy scale for low-lying excitations in the system. Using exact diagonalization for small system sizes and DMRG for larger chains (up to $L=200$)~\cite{white1992,jesenko2011finite}, we found that both systems have finite energy gaps, equal to $\Delta_1 = 8.87$ and $\Delta_2 = 0.51$ respectively. We deliberately choose the parameters of the models to yield energy gaps at different scales, while maintaining the microscopic interaction energy $J_z$ on the same order for both models. This allows us to study the interplay between different energy scales and the effectiveness of our driving scheme at low temperatures.
\section{Methods}
\label{sec:methods}

In this section, we describe the non-equilibrium setup used to study energy transport, the tensor network techniques used to find the non-equilibrium steady state, and the procedure for determining the local temperature of the system out of equilibrium.

\subsection{Setup}
\label{sec:setup}

\begin{figure}
\begin{center}
\includegraphics[width=\columnwidth]{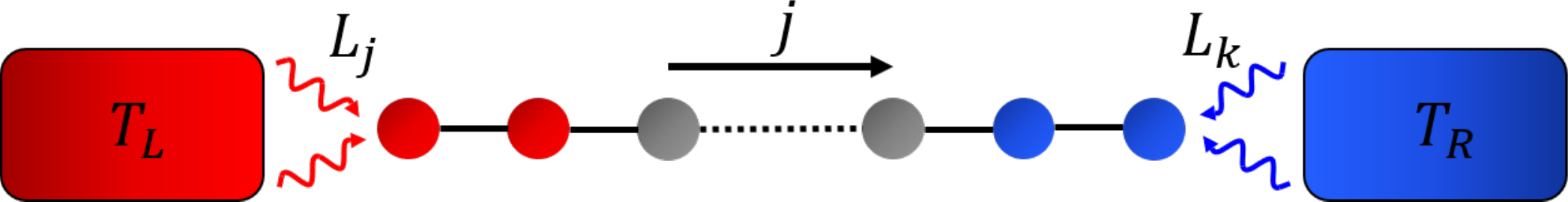}
\caption{Schematic diagram of our non-equilibrium setup. The spin chain is connected at both ends to thermal baths at temperatures $T_L$ and $T_R$. Each bath acts on two boundary spins via Lindblad operators $L_{k}$. In NESS, a homogeneous current $j$ flows through the bulk.}
\label{fig:config}
\end{center}
\end{figure}

Our non-equilibrium configuration is shown in \figref{fig:config}. The setup consists of a one-dimensional spin chain coupled to two baths at its ends. The bath dynamics is chosen such that, when decoupled from the rest of the chain, the left and right leads are maintained at temperatures $T_{L, R} = T_b \pm \delta T$, where $T_b$ is the average bath temperature and $\delta T$ is a small temperature imbalance meant to drive the system out of equilibrium. Once the coupling to the chain is turned on, the baths are still continuously driven towards their decoupled thermal state, but the coupling with the system causes the steady state of the bath to differ from its target thermal state. Nevertheless, the baths impose an energy gradient in the spin chain and we are able to accurately measure the true effective temperature. In practice, in order to obtain well-differentiated energy profiles even at low temperatures, we set $\delta T = 0.2T_b$. The resulting NESS carries an energy current $j$, which is uniform in the bulk of the chain. 

The evolution of the system coupled to an environment can be modeled using Markovian dissipative dynamics. The time evolution of the density matrix of the system is governed by Lindblad's equation~\cite{lindblad1976generators,gorini1976}

\begin{equation}
    \frac{d\rho}{dt} = \mathcal{L}(\rho) \equiv -i[H, \rho] + \sum_k\Big(L_k\rho L_k^\dagger - \frac{1}{2}\{L_k^\dagger L_k, \rho\}\Big), 
    \label{eq:lindblad}
\end{equation}
where $H$ is the full system Hamiltonian describing the coherent bulk dynamics and $L_k$ are Lindblad operators describing the coupling to the baths. The operator $\mathcal{L}$ is called the Liouvillean super-operator acting on the space of density matrices. We set $\hbar=1$ throughout this paper. 

In our case, we have two sets of dissipators $L_k$ corresponding to the left and right baths. Each Lindblad operator acts either on the two rightmost or leftmost sites of the chain. Note that unlike in the case of spin transport, a two-site bath is required here in order to have an efficient coupling to the energy density and induce a finite-temperature thermal state at the boundaries~\cite{prosen2009matrix,mendoza2015}. The transport properties in the bulk should not depend on the specific implementation of the baths, as long as the dynamics is sufficiently ergodic. A detailed description of the Lindblad operators is provided in \appref{sec:appendixA}. 

\subsection{Tensor Network Approach}
\label{sec:TN}

The combination of coherent and dissipative dynamics leads to a non-trivial NESS $\rho_\infty$, which is the fixed point of the Lindblad equation $d\rho_\infty / dt = 0$.  Formally, the solution to the Lindblad equation is given by $\rho(t) = e^{\mathcal{L}t}\rho(0)$ and the NESS corresponds to the infinite time limit $\rho_\infty = \lim\limits_{t\rightarrow\infty}\rho(t)$ when all the observables have reached their stationary values. However, since the size of the operator space grows exponentially as $4^L$ for a spin-1/2 chain, it becomes very challenging to find steady state solutions for large $L$. While there exist exact solutions for the NESS of non-interacting systems~\cite{Prosen2008,zanoci2016entanglement} and certain strongly-driven interacting systems~\cite{clark2010exact,karevski2013exact,prosen2011exact,prosen2014exact,popkov2020exact}, one usually has to resort to approximate tensor network methods when dealing with generic interacting systems and arbitrary driving~\cite{vidal2004efficient,verstraete2004matrix,zwolak2004,cui2015variational,mascarenhas2015matrix}. We will use a tensor network representation of the density matrix and perform an approximate time evolution under the Liouvillean super-operator using the Time Evolving Block Decimation (TEBD) algorithm~\cite{vidal2003efficient,vidal2004efficient,schollwock2011density}. 

First, we map the MPO representation of the density matrix onto an equivalent MPS form with a local Hilbert space dimension of $4$. To achieve this, we vectorize the density operator $\rho$ by reshaping it into a column vector $\ket{\rho}$ ~\cite{verstraete2004matrix,zwolak2004,mascarenhas2015matrix}. Next, we express the Liouvillean in this new vectorized form, using the fact that $\ket{X\rho Y}=Y^T\otimes X\ket{\rho}$ for arbitrary matrices $X$ and $Y$~\cite{mascarenhas2015matrix}, and obtain

\begin{equation}
\begin{split}
    \mathcal{L} = &-i(\mathbb{1}\otimes H - H^T\otimes\mathbb{1}) \\
    &+ \sum_k\Big(L_k^*\otimes L_k - \frac{1}{2}(\mathbb{1}\otimes L_k^\dagger L_k + L_k^TL_k^*\otimes\mathbb{1})\Big).
\end{split}
\end{equation}
The time-evolution operator $e^{\mathcal{L}t}$ can now be Trotterized and applied to $\ket{\rho}$ using the standard TEBD approach~\cite{verstraete2004matrix,zwolak2004}. We use a second-order Suzuki-Trotter decomposition with a time step of $\delta t=0.05$~\cite{suzuki1990fractal}, which is small enough so as to not dominate over the truncation error. The NESS $\rho_\infty$ is approximated by evolving an initial state for a sufficiently long time $t=1000$. More details on our numerical implementations are available in \appref{sec:appendixB}. 

In our numerical approach, there are implicit assumptions that the NESS solution is unique (independent of the initial state $\rho(0)$) and only depends on the thermodynamic parameters of the baths (in this case, the temperatures $T_{L, R}$). These assumptions are justified for non-integrable chaotic systems possessing good thermalization properties~\cite{prosen2009matrix,bertini2020finite}. The convergence time to this NESS solution is given by the inverse gap of the Liouvillean super-operator $\mathcal{L}$~\cite{Prosen2008,prosen2009matrix}. At low temperatures, the convergence can be very slow and we were not able to reach a solution within reasonable time for bath temperatures below $T_b=1$. This sets the limit of our minimum accessible temperature, as we will see in \secref{sec:thermal_results}.

\subsection{Local Temperature}
\label{sec:thermal}

After finding the NESS of a boundary-driven quantum system, one can ask whether this state can be described by local thermal equilibrium. If this is indeed the case, then a local temperature $T$, which may differ considerably from the bath temperature $T_b$, can be assigned. Determining an effective temperature for the system is crucial in obtaining the correct functional dependence of the transport coefficients. 

The locality of temperature is a challenging issue even in equilibrium quantum systems~\cite{hartmann2004,hartmann2005,hartmann2006minimal,garcia2009,ferraro2012intensive,kliesch2014locality}. The problem is that a subsystem of a thermal state is generally not in a state with a locally well-defined temperature. The interactions between the subsystem and its environment create correlations that can lead to significant deviations from a thermal state. A local temperature can be assigned unambiguously only if these correlations decay exponentially with distance~\cite{kliesch2014locality}.

For closed quantum systems in non-equilibrium, it has been shown that local density matrices relax to a thermal state only if the Hamiltonian is non-integrable~\cite{rigol2007,rigol2008thermalization,rigol2009,gogolin2011,steinigeweg2013,gogolin2016equilibration}. It is believed that this is achieved by means of eigenstate thermalization~\cite{deutsch1991,srednicki1994}, although other mechanisms have also been proposed~\cite{rigol2012,sirker2014,steinigeweg2014}. The thermalization properties of open driven systems have been less studied.

In this paper, we propose a method for inferring the local temperature based on the assumption that the system is in local thermal equilibrium in the steady state. This procedure can be understood as a gradient expansion in the spirit of hydrodynamics. Let $\tilde{\rho}_A$ denote the NESS reduced density matrix of a small set of sites $A$ and let $\rho_A(T)$ denote the local density matrix of the same set of sites in a uniform thermal state. Then we assign the local temperature of $A$ to be the temperature $T_A$ which minimizes the trace distance between the two density matrices

\begin{equation}
   D(\tilde{\rho}_A, \rho_A(T)) = \frac{1}{2}\Tr\left(\sqrt{\left(\tilde{\rho}_A-\rho_A(T)\right)^2} \right).
   \label{eq:trace_dist}
\end{equation}
The trace distance between the two states is a useful metric because it places an upper bound on the difference between the corresponding expectation values of any local observable~\cite{mendoza2015}. However, we still have to compare these expectation values in the two states to conclude that $\tilde{\rho}_A$ is actually well approximated by $\rho_A(T_A)$. This procedure is guaranteed to give the correct temperature when all gradients vanish, i.e. to zeroth order in the gradient expansion. Note that the precise assignment can depend on the size of $A$ and the size of the total system used to define the global thermal state. Beyond the zeroth order gradient expansion, the concept of local temperature becomes ambiguous and a consistent scheme for incorporating higher order corrections has to be specified. Therefore, we restrict ourselves here to the zeroth order expansion.

In practice, we choose the region $A$ to be comprised of pairs of consecutive sites $(i, i+1)$ in the chain, which is consistent with our definition of local energy. To produce an exact thermal state, we take a smaller version of our system (i.e. same Hamiltonian, without coupling to baths) with only $L=10$ sites and compute its equilibrium thermal state via matrix exponentiation for a range of temperatures $T$. Using these global thermal states, we evaluate the reduced density matrix on the two middle sites $\rho_2(T)$. The computational complexity limits us to rather small systems. However, we verified that the finite size effects decrease rapidly with $L$ and become negligible around $L\sim10$, as long as we compute our reduced density matrix $\rho_2(T)$ away from the boundaries. In our numerical implementation, we choose a set of temperatures in the range $1\leq T\leq 100$ in increments of $0.01$. This allows for a great resolution when determining the local temperature $T_{i, i+1}$ using \eqref{eq:trace_dist}. Our optimization procedure is a simple grid search over the aforementioned temperature range.

Notice that we do not rely on a particular ansatz for the local density matrix $\rho_2(T)$, which is not a thermal state, but rather the state of a sub-system of a thermal state. This is in contrast with the approach introduced in Ref.~\cite{mendoza2015}, where the authors extended the concept of local thermal states to driven open-system
non-equilibrium setups. They showed that local thermalization arises in non-integrable thermally-driven systems, which is exactly the case for our models, thus further justifying our assumption of local thermal equilibrium in NESS. Their approach can be used in conjunction with our method to reveal additional information about the structure of the local state $\rho_2(T)$, as explained in \appref{sec:appendixC}. In \secref{sec:thermal_results} we will show that the two methods agree very well, both in terms of the local temperature and the expectation values of local observables. 

\section{Results}
\label{sec:results}

We perform numerical simulations of the non-equilibrium dynamics for a range of driving temperatures $1\leq T_{b}\leq 40$. The system size is set to $L=100$, unless specified otherwise. When computing expectation values of various observables, approximately $5$ sites were discarded at each boundary to remove the effects of driving. To emphasize that our methods are applicable to a wide range of temperatures, we display results for high ($T_b=20$), intermediate ($T_b=5$), and low ($T_b=1.5$) driving temperatures. In order to improve the readability of the figures, in most cases we only show results for the XZ model, with the understanding that similar results hold for the Ising model in a tilted field as well. 

\subsection{Diffusive Transport}
\label{sec:diffusion_results}

\begin{figure}[ht]
\begin{center}
\includegraphics[width=\columnwidth]{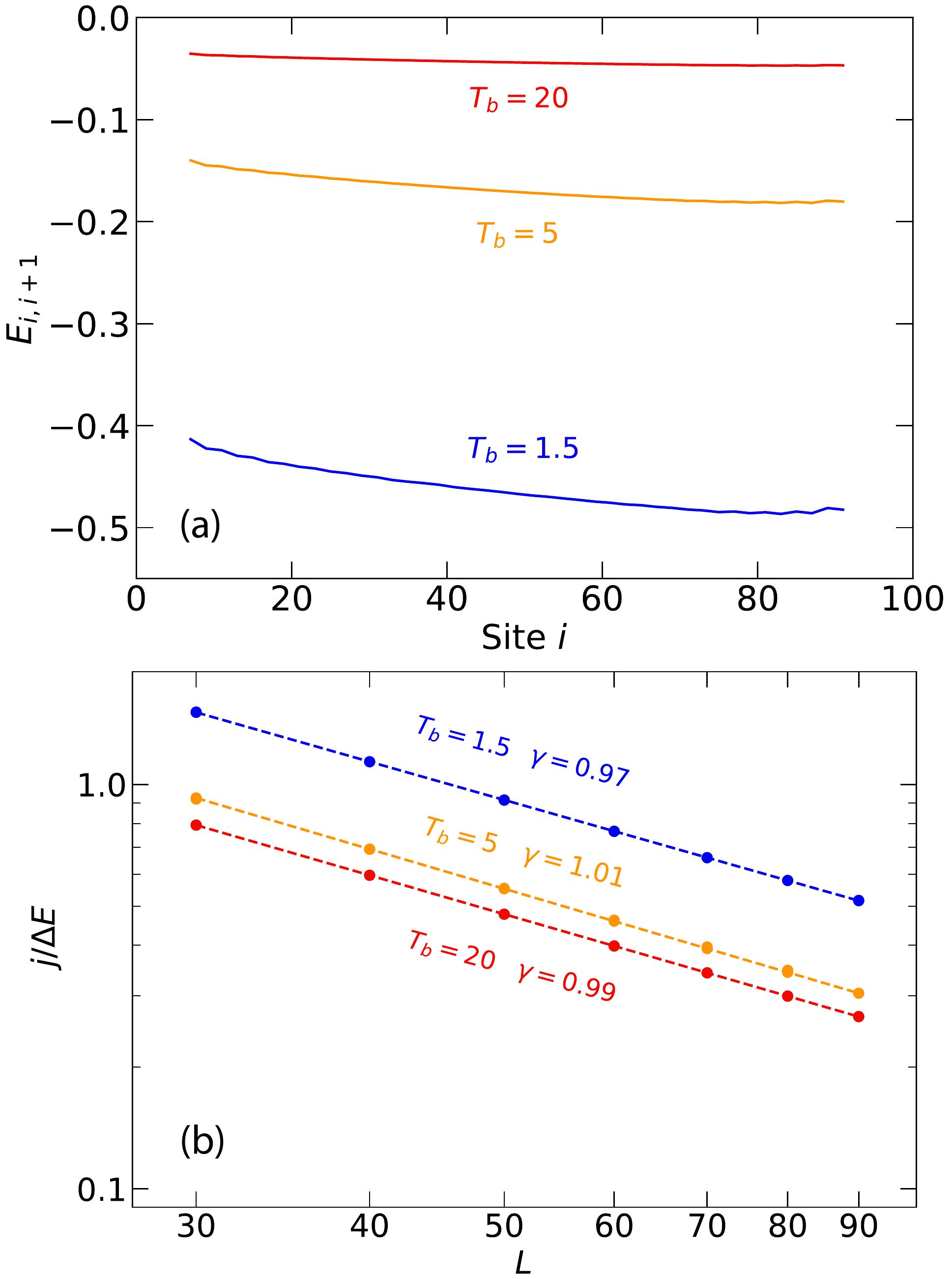}
\caption{NESS transport properties of the XZ model at different bath temperatures $T_b$. (a) Spatial energy profiles for a spin chain of length $L=100$, showing a constant gradient $\nabla E$. (b) Scaled energy current $j/\Delta E$ as a function of system size. Symbols are numerical values and lines represent fits to the scaling $j/\Delta E=-D/L^\gamma$. These values of $\gamma$ indicate diffusion according to Fourier's law. Analogous results for the Ising model in a tilted field can be found in Ref.~\cite{prosen2009matrix}.}
\label{fig:ness}
\end{center}
\end{figure}

We begin our investigation of the thermally-driven systems by exploring the nature of energy transport through our two models (see \eqref{eq:model_1} and \eqref{eq:model_2}). In \figref{fig:ness}(a) we plot the on-bond energy $E_{i, i+1}=\langle H_{i, i+1}\rangle$ as a function of lattice site $i$. At lower temperatures, the edge effects become more prominent. Nonetheless, the energy profiles are linear in the bulk, suggesting diffusive energy transport. In addition, the current scales linearly with inverse system size, as shown in \figref{fig:ness}(b). We find that the scaling exponent $\gamma$ is close to $1$ for all temperatures, which corresponds to diffusive transport. Our results are consistent with previous studies on energy transport in the same models~\cite{prosen2009matrix,ye2019emergent}. 

\subsection{Local Temperature}
\label{sec:thermal_results}

\begin{figure}[ht]
\includegraphics[width=\columnwidth]{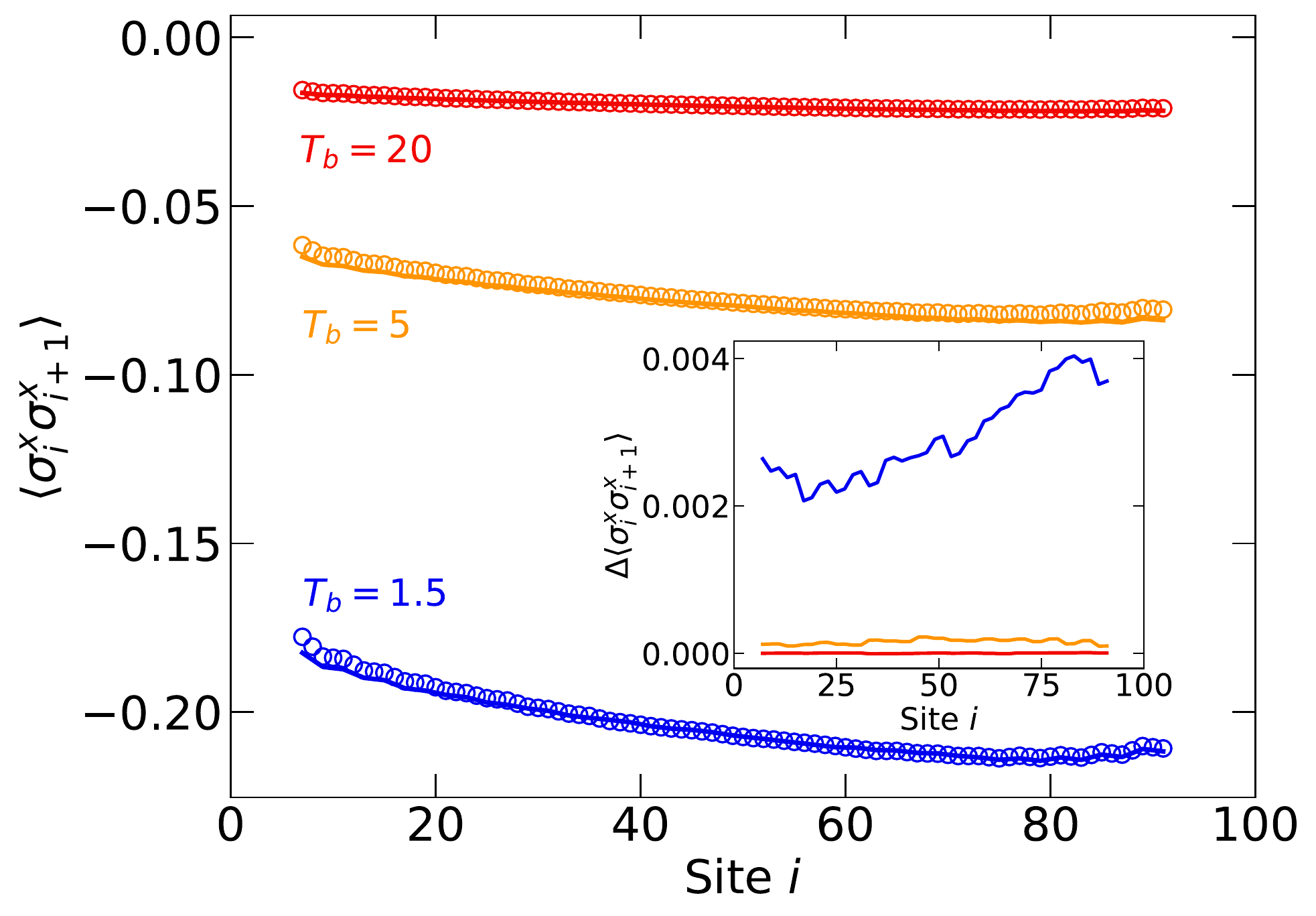}
\caption{Comparison between local expectation values in the XZ chain at different bath temperatures $T_b$. The two-point functions $\langle \sigma_i^x \sigma_{i+1}^x \rangle$ measured directly in NESS (circles) match with the ones computed using the reduced density matrix $\rho_2$ of a thermal state (solid lines). Inset shows the difference in expectation values between our method introduced in \secref{sec:thermal} and the one described in \appref{sec:appendixC}. Similar results also hold for the tilted-field Ising model.}
\label{fig:exp_vals}
\end{figure}

\begin{figure}[ht]
\begin{center}
\includegraphics[width=\columnwidth]{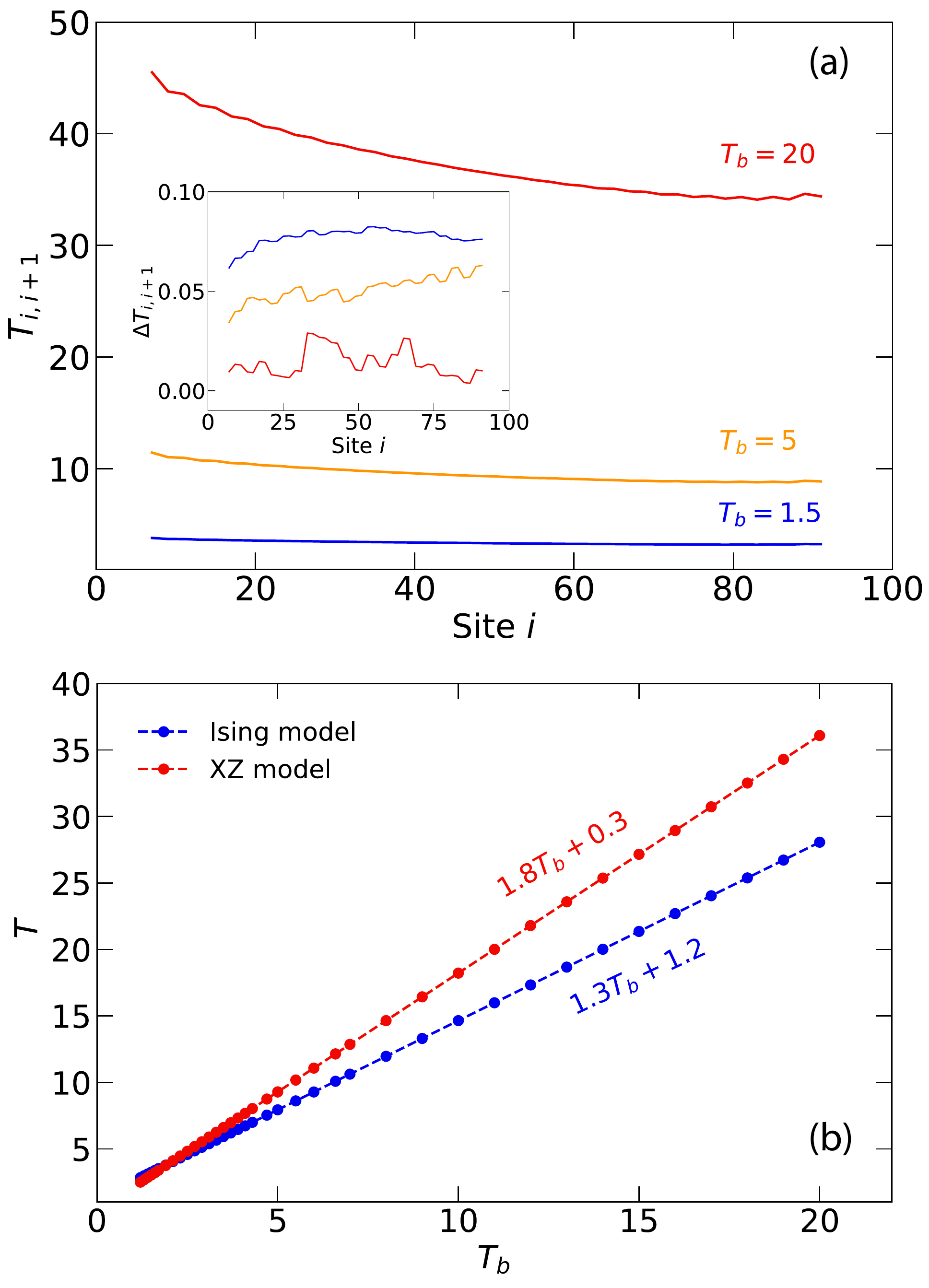}
\caption{Local temperature in NESS of the XZ model at different bath temperatures $T_b$. (a) Spatial temperature profiles for a spin chain of length $L=100$ and the same NESS as in \figref{fig:ness}(a). Inset shows the difference in the local temperatures extracted via our two methods. (b) Local temperature $T$ in the middle of the chain as a function of bath temperature $T_b$ for the two models. Symbols are numerical values and lines represent linear fits.}
\label{fig:local_temp}
\end{center}
\end{figure}

Next, we examine the local thermalization properties of the non-equilibrium states using the two methods described in \secref{sec:thermal} and \appref{sec:appendixC}. For the first method, we identify the temperature of the thermal system whose local density matrix is closest to $\tilde{\rho}_{i, i+1}$ in NESS for each pair of neighboring spins in the bulk. Similarly, for the second method, we find the local thermal state $\rho_{i, i+1}$ that best matches $\tilde{\rho}_{i, i+1}$. In both cases we find good solutions with trace distances ranging between $10^{-4}$ at high temperatures and $10^{-3}$ at low temperatures.

To confirm that the solutions indeed capture the local properties of the non-equilibrium states, we compare the corresponding expectation values for one-point $\langle\sigma_i^\alpha\rangle$ and two-point functions $\langle\sigma_i^\alpha\sigma_{i+1}^\alpha\rangle$. In \figref{fig:exp_vals} we plot the results for two-point functions and $\alpha=x$, but similar conclusions can be drawn for the one-point functions and other Pauli matrices. The deviations in expectation values range from $2\%$ to $4\%$ for the one-point functions and from $3\%$ to $5\%$ for the two-point functions, with better agreement at high temperatures. These numbers are on par with previous results~\cite{mendoza2015}. The distinction between the two methods is only noticeable at the lowest temperatures (see inset of \figref{fig:exp_vals}), and even then the discrepancy is less than $2\%$. We can thus conclude that both methods result in states that capture very well the local properties of NESS. Moreover, since the two methods agree, we can infer that the local states are well described by the ansatz in \eqref{eq:rho_thermal}.

Once we have checked the validity of our solutions, we plot the local temperature profiles for the driven states in \figref{fig:local_temp}(a). Away from the edges, the temperature varies uniformly from left to right. We also show the difference in temperature $\Delta T_{i, i+1}$ when comparing our two methods. This difference is negligible for all practical purposes, once again confirming the agreement between the two techniques. The temperature profiles can be used to extract the thermal conductivities and heat capacities via an Einstein relation. We discuss this further in \appref{sec:appendixD}.

Finally, we find that the local temperature $T$ in the middle of the chain scales linearly with the bath temperature $T_b$, as can be seen in \figref{fig:local_temp}(b). The local temperature is much larger than the average driving temperature for both models. This is consistent with previous studies where only the energy density was used to infer the temperature~\cite{znidaric2010,vznidarivc2011transport}. Therefore, if we had used $T_b$ as our non-equilibrium temperature, we would have obtained a completely different scaling of the diffusion constant with temperature. This justifies our attempts to assign a local temperature in a more systematic way. 

As we mentioned before, $T_b=1$ is the smallest bath temperature for which we were still able to reliably converge to a NESS solution. This corresponds to a minimum effective temperature of $T=2.5$ for the Ising model and $T=2.1$ for the XZ model. We further discuss the limiting factors for this temperature in \secref{sec:discussion}.

\subsection{Temperature Dependence of Diffusion Constants}
\label{sec:temp_dep_results}

\begin{figure}[ht]
\begin{center}
\includegraphics[width=\columnwidth]{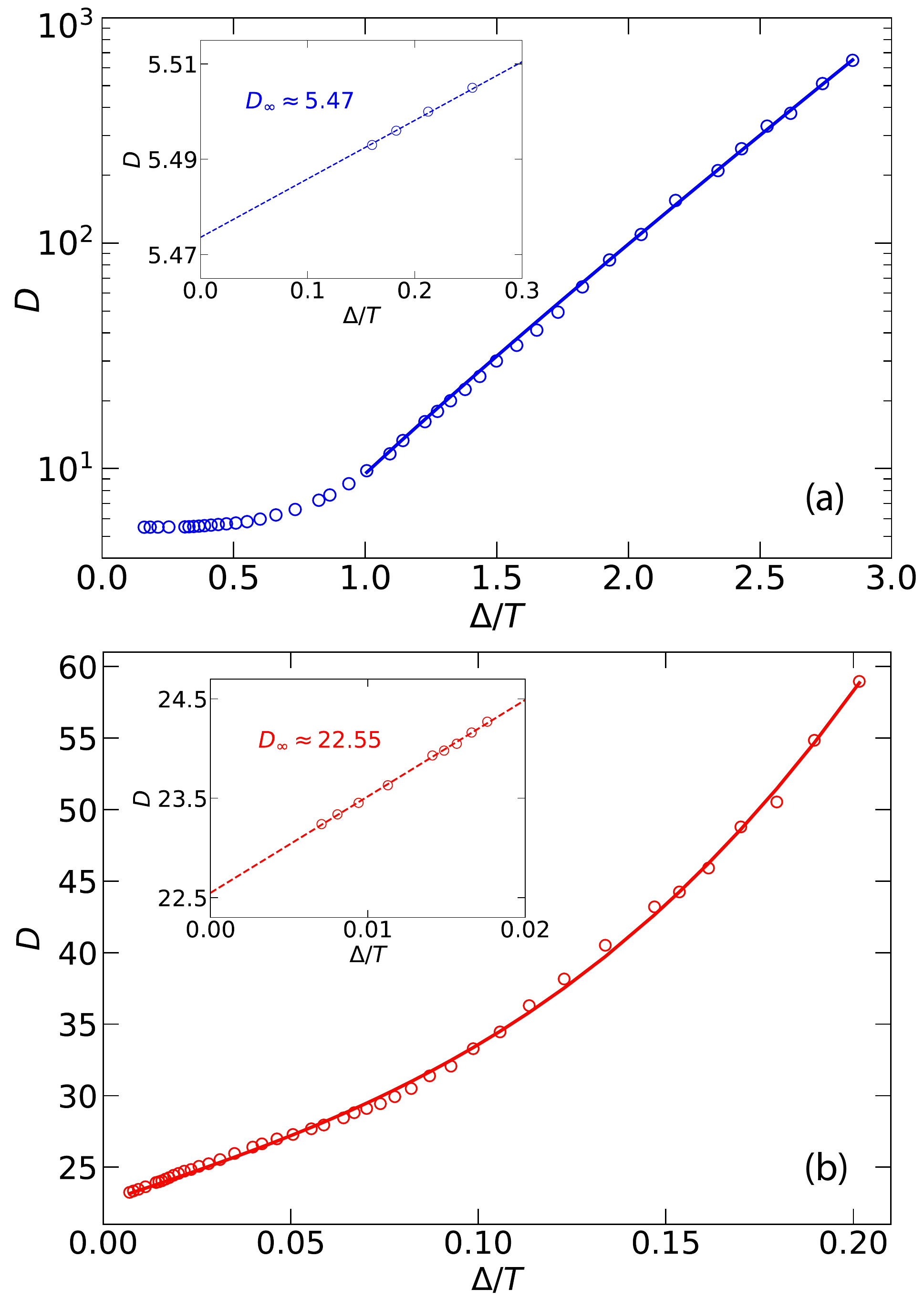}
\caption{Temperature dependence of the energy diffusion constants $D$ for (a) tilted-field Ising model and (b) XZ model. Symbols represent numerical values and solid lines are fits to \eqref{eq:D_semi_class} and \eqref{eq:D_series} with $k=3$ respectively. Insets show linear extrapolations to infinite temperature. All inverse temperatures are scaled by the corresponding energy gap $\Delta$.}
\label{fig:diff_const}
\end{center}
\end{figure}

We now combine the results of the previous sections to obtain the temperature dependence of our diffusion constants, as shown in \figref{fig:diff_const}. In the high-temperature limit, $D(T)$ approaches a constant (temperature-independent) value $D_\infty$. We can find this value by extrapolating to infinite temperature (see insets of \figref{fig:diff_const}). Once we decrease the temperature, the diffusion constant becomes larger. For gapped systems at low temperatures, the energy gap $\Delta$ is the relevant energy scale for transport. The precise functional form of $D(T)$ depends on the different temperature regimes relative to $\Delta$, which we now discuss. 

At very low temperatures $T\ll \Delta$ we have a diluted gas of excitations, whose kinetic properties determine the transport coefficients~\cite{damle1998spin,damle2005universal}. The dispersion relation for these particles is given by a low-momentum expansion

\begin{equation}
    \epsilon_k = \Delta + \frac{c^2k^2}{2\Delta} + \mathcal{O}(k^4),
\end{equation}
where $c$ is a velocity related to the excitation mass $\Delta/c^2$. The excitations can be treated semi-classically using the Maxwell-Boltzmann statistics $n_k = e^{-\beta\epsilon_k}$. Their concentration is given by

\begin{equation}
    n = \frac{1}{L}\int \frac{dk}{2\pi}n_k \sim \sqrt{T}e^{-\Delta/T},
\end{equation}
where we set $k_B=1$ and only kept track of the temperature-dependent part. The characteristic thermal velocity of these particles is $v=c\sqrt{T/\Delta}$. In 1D, the energy and momentum conservation laws make it impossible to have energy changes as a result of two-body collisions. Therefore, three-body collisions must be the primary scattering process for energy transport. From dimensional analysis, the diffusion constant scales as $D\sim v^2\tau \sim v\ell$, where $\tau$ is the average time between collisions and $\ell$ is the mean free path. We can assume, for simplicity, that a three-body collisions occurs whenever the particles are within a distance $r$ of each other. From standard kinetic theory we know that the mean free path between two-particle collisions is $(n\sigma)^{-1}$, where $\sigma$ is an effective cross-section. For three-body collisions, we factor in the probability $nr$ of having a third particle nearby when a two-body collision occurs and obtain the scaling $\ell\sim (n^2\sigma r)^{-1}$. Assuming that $\sigma$ and $r$ are temperature-independent, we finally obtain the low-temperature ($T\ll \Delta$) dependence

\begin{equation}
    D \sim \frac{v}{n^2} \sim \frac{e^{2\Delta/T}}{\sqrt{T}}.
    \label{eq:D_semi_class}
\end{equation}

This semi-classical prediction suggests an exponential increase in the diffusion constant at low temperatures. Note that similar exponential scalings can be derived for transport coefficients of other conserved quantities~\cite{damle1998spin,damle2005universal}. For example, spin transport only requires two-body collisions and the corresponding diffusion constant scales as $D_{\text{spin}}\sim v/n\sim e^{\Delta/T}$.

Going from low to intermediate ($T\gtrsim \Delta$) and high temperatures ($T\gg \Delta$), the semi-classical description is no longer applicable. Instead, we perform a power-series expansion around the high-temperature result:

\begin{equation}
    D = D_\infty \left(1+\sum_{k\geq 1}\frac{c_k}{T^k}\right),
    \label{eq:D_series}
\end{equation}
where $c_k$ are some constants depending on the microscopic parameters of the model. 

Although we drive the chains for both Hamiltonians to approximately the same lowest temperature, the tilted-field Ising model has a much larger energy gap and hence it actually reaches a regime where $T\ll \Delta$. For these temperatures, the diffusion constant matches the semi-classical prediction in \eqref{eq:D_semi_class} remarkably well, as can be seen in \figref{fig:diff_const}(a). We emphasize that $\Delta$ is not a fitting parameter, but rather the exact numerical value specified in \secref{sec:models}. We only fit the overall proportionality factor in \eqref{eq:D_semi_class}. On the other hand, the XZ model has a small energy gap and we only managed to cool the system down to $T\approx 5\Delta$. From \figref{fig:diff_const}(b) we see that $D(T)$ increases linearly (with inverse temperature) at high temperatures up until roughly $T\approx 10\Delta$, where higher order corrections become significant. We find that the inverse-temperature series expansion in \eqref{eq:D_series} with $k=3$ is sufficient to capture the different regions of our data. If we could further reduce the temperature below the energy gap, we would expect to find the same exponential increase in the diffusion constant as for the tilted-field Ising model. 

Also notice that the XZ model has a much larger diffusion constant compared to the tilted-field Ising model at the same scaled inverse temperature $\Delta/T$. This suggests the presence of an extra scale in the model, which is different from the interaction strength or the energy gap. One possibility is that the model is close to an integrable point.
\section{Discussion}
\label{sec:discussion}

In this work, we studied numerically the non-equilibrium steady states of two non-integrable 1D models in a boundary-driven setup for a wide range of temperatures. We showed that both models feature diffusive energy transport, even at low temperatures. We also showed that the resulting NESS is well described by local thermal states and proposed a new method for computing its local temperature. Based on this method, we found that the local temperatures in the bulk of the chains are much higher than the driving bath temperatures.  

Subsequently, we extracted the temperature dependence of the energy diffusion constants. We showed that for the tilted-field Ising model at low temperature ($T\ll \Delta$), the diffusion constant scales exponentially with inverse temperature, in agreement with the semi-classical predictions for gapped one-dimensional systems~\cite{damle1998spin,damle2005universal}. Unfortunately, for the XZ model with a much smaller gap $\Delta$, we were unable to reach temperatures below the energy gap with our current approach. However, for the available regime of temperatures, we showed that the diffusion constant increases polynomially with inverse temperature. 


We would like to draw a comparison between the two models studied in this paper. Using the two-site baths, we were able to cool the chain to temperatures much lower than the gap $\Delta$ for the Ising model, but not for the XZ model. This suggests that the energy gap is not the limiting factor when considering open-system dynamics. The minimum accessible temperature is determined by the convergence rate to NESS, which in turn depends on the spectral gap of the Liouvillean super-operator $\mathcal{L}$~\cite{Prosen2008,prosen2009matrix}. Therefore, as long as both models have their microscopic parameters on the same scale, it is plausible that both Liouvilleans would have similar spectral gaps and we would be able to reach comparable minimum temperatures. This suggests a fundamental limitation on the efficiency of the two-site driving scheme which is independent of the model under study. It is very likely that at low temperatures, the energy levels of the bath simply do not couple well to those in the bulk, resulting in poor driving. In our previous work~\cite{zanoci2016entanglement}, we have showed that, at least for non-interacting fermions, a much larger bath is required to efficiently cool the system to very low temperatures. Devising a general framework for constructing efficient baths capable of driving more general interacting models to lower temperatures is an important direction of future research.

One possible extension of our analysis involves transport in gapless non-integrable spin systems. An ubiquitous example of such a system is the XXZ model in a staggered magnetic field. For certain values of the anisotropy, the perturbation due to the magnetic field is irrelevant and the model at low temperatures can be described by a gapless Luttinger liquid~\cite{giamarchi2003quantum}. Transport for this model is no longer expected to be diffusive~\cite{Bulchandani12713}.  Previous studies based on dynamical typicality and finite-temperature t-DMRG indicate that electrical and thermal conductivities exhibit power-law dependencies at low temperature, with exponents that are functions of the Luttinger parameter $K$~\cite{huang2013,steinigewegs2015}. It would be interesting to confirm this scaling using our open-system framework. However, the regime of temperatures in which this power-law behavior manifests itself may be too low for our current approach and further improvements to the driving scheme may be required first.    

Experimental measurements of the temperature-dependent diffusion constants could be performed in the near future. Experiments with ultracold atoms in optical lattices provide a promising route for investigating out-of-equilibrium quantum many-body systems. Recently, several groups have reported charge, spin, and heat diffusion in the two-dimensional Fermi-Hubbard model using quantum-gas microscopes~\cite{brown2019bad,nichols2019spin,guardado2020subdiffusion}. This 2D system can be further divided into separate 1D systems, where various transport phenomena, such as spin-charge separation, can be observed~\cite{Salomon2019,vijayan186}. Additionally, a simulation of Heisenberg-type models can be achieved in the strong-coupling limit of the Fermi-Hubbard model or using a two-component Bose-Hubbard model. The latter was recently used to study spin transport far from equilibrium after quantum quenches in a 1D Heisenberg XXZ model~\cite{jepsen2020spin}. Energy diffusion could be studied in a similar way, although it is somewhat more challenging because it requires local temperature measurements. 

The aforementioned experiments typically implement closed quantum systems, which is a different setup than the one considered in our work. A more related architecture involves coupling two macroscopic cold atom reservoirs via a low-dimensional mesoscopic channel~\cite{Krinner2017}. Particle and energy transport, induced either by a chemical potential or temperature difference respectively, has been previously studied~\cite{brantut2012conduction,Stadler2012,brantut2013thermoelectric,lebrat2018}. One could potentially apply these experimental techniques to observe an exponential increase in the diffusion constant at low temperatures, as predicted by our numerical results.  It would also be interesting to adapt our methods to study NESS on quantum devices, perhaps by physically implementing the boundary driving using an appropriate quantum circuit with ancilla qubits.

\begin{acknowledgments}
We are grateful to Christopher White and Subir Sachdev for valuable discussions. CZ acknowledges financial support from the Harvard-MIT Center for Ultracold Atoms through the NSF grant no. PHY-1734011. BS acknowledges support from the U.S. Department of Energy, Office of Science, Office of Advanced Scientific Computing Research, Accelerated Research for Quantum Computing program.
\end{acknowledgments}

\bibliographystyle{apsrev4-2}
\bibliography{references}

\appendix
\section{Two-site Lindblad Operators}
\label{sec:appendixA}

In order to induce a NESS at finite temperature, we use the so-called two-site Lindblad operators~\cite{prosen2009matrix,vznidarivc2011transport,mendoza2015,mendoza2019asymmetry}, which act on two boundary spins at each end. We want to construct a super-operator $\mathcal{L}_B$ from a set of Lindblad operators $\{L_k\}$ such that it drives the two sites to a Gibbs state at some temperature $T$, i.e. $\mathcal{L}_B(\rho_B)=0$ where 

\begin{equation}
    \rho_B = \frac{e^{-h/T}}{\Tr\big(e^{-h/T}\big)}
\end{equation}
and $h$ is the Hamiltonian for the two spins ($h=H_{1, 2}$ for the left bath and $h=H_{L-1, L}$ for the right bath). Therefore we require that $\rho_B$ is a unique eigenvector of $\mathcal{L}_B$ with eigenvalue $0$. In addition, we will impose that all the other eigenvalues are equal to $-1$, which leads to the fastest convergence to $\rho_B$~\cite{prosen2009matrix}. Our construction is equivalent to the one described in Refs.~\cite{prosen2009matrix,mendoza2015}, but is formulated in a slightly different way.  

First, we diagonalize the density matrix $\rho_B = V^\dagger d V$, where $d=\text{diag}(d_0, d_1, d_2, d_3)$ and $V$ is unitary. Then it is easy to directly check that the following set of $16$ operators satisfy the requirements above:

\begin{equation}
    \tilde{L}_{ij} = \sqrt{\Gamma d_m} r_i\otimes r_j, \quad m = (j \bmod 2) + 2(i \bmod 2),
\end{equation}
where $i, j = 0, 1, 2, 3$, $r_{0, 1} = \frac{1}{2}(\sigma^x\pm i\sigma^y)$ and $r_{2, 3} = \frac{1}{2}(\sigma^0\pm \sigma^z)$. Here $\Gamma$ quantifies the overall strength of the bath coupling and we choose $\Gamma = 1$. Notice that there is a minor difference in the index labeling compared to Ref.~\cite{prosen2009matrix}.

Now that we have 

\begin{equation}
    \tilde{\mathcal{L}}_B(d) = \sum_{i, j=0}^3 \big(\tilde{L}_{ij}d \tilde{L}_{ij}^\dagger - \frac{1}{2}\tilde{L}_{ij}^\dagger \tilde{L}_{ij}d - \frac{1}{2}d\tilde{L}_{ij}^\dagger \tilde{L}_{ij}\big) = 0,
\end{equation}
we can multiply it by $V^\dagger$ and $V$ on the left and right sides, and use the identity $VV^\dagger = \mathbb{1}$ to deduce that

\begin{equation}
\begin{split}
    \sum_{i, j=0}^3 &\left(V^\dagger\tilde{L}_{ij}d \tilde{L}_{ij}^\dagger V - \frac{1}{2}V^\dagger\tilde{L}_{ij}^\dagger \tilde{L}_{ij}dV - \frac{1}{2}V^\dagger d\tilde{L}_{ij}^\dagger \tilde{L}_{ij}V\right)\\
    &= \sum_{i, j=0}^3 \left(L_{ij}\rho_B L_{ij}^\dagger - \frac{1}{2}L_{ij}^\dagger L_{ij}\rho_b - \frac{1}{2}\rho_BL_{ij}^\dagger L_{ij}\right) \\
    &= \mathcal{L}_B(\rho_B) = 0,
\end{split}
\end{equation}
with $L_{ij} = V^\dagger \tilde{L}_{ij}V$. Hence we can use these new Lindblad operators $L_{ij}$ to construct $\mathcal{L}_B$. Moreover, since $V$ is unitary, the eigenvalues of $\mathcal{L}_B$ will be the same as those of $\tilde{\mathcal{L}}_B$, as desired.  
\section{Numerical Implementation Details}
\label{sec:appendixB}

\begin{figure}[t]
\begin{center}
\includegraphics[width=\columnwidth]{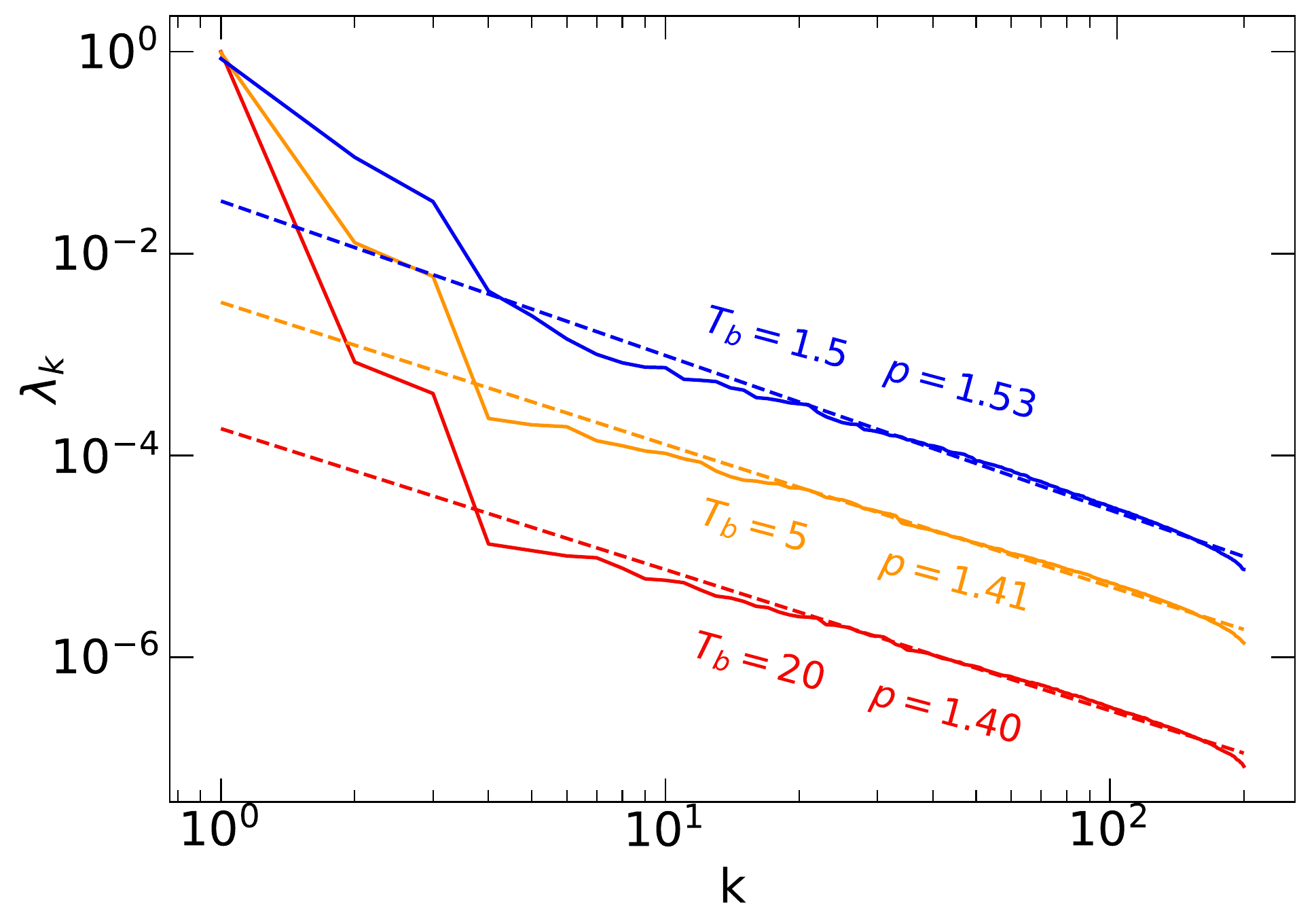}
\caption{Schmidt spectrum for the NESS of a boundary driven XZ chain of size $L=100$ with a bond dimension $\chi=200$ at different bath temperatures $T_b$. Dashed lines represent fits to a power law decay $\lambda_k \sim k^{-p}$. Similar results for the Ising model in a tilted field can be found in Ref.~\cite{prosen2009matrix}.}
\label{fig:schmidt}
\end{center}
\end{figure}

As mentioned in the main text, we simulate the time evolution of the vectorized density matrix using the TEBD algorithm~\cite{vidal2003efficient,vidal2004efficient,verstraete2004matrix,zwolak2004}. During the evolution, we restrict the amount of built-up entanglement by truncating the matrix products to a maximum bond dimension $\chi$. The size of the truncation error is related to the operator space entanglement entropy of the NESS~\cite{prosen2007}. For a bipartite splitting of the chain into two regions, A and B, we can write 

\begin{equation}
   \ket{\rho} = \sum_k \sqrt{\lambda_k}\ket{\xi_k^A}\ket{\xi_k^B},
\end{equation}
where $\ket{\xi_k^{A, B}}$ are orthogonal vectors and the Schmidt coefficients $\sqrt{\lambda_k}$ satisfy $\sum_k \lambda_k = 1$. If we only keep the first $\chi$ non-zero Schmidt coefficients, the truncation error is equal to $\sum_{k>\chi}\lambda_k$. For boundary-driven dynamics, one typically observes an asymptotic decay $\lambda_k\sim k^{-p}$~\cite{bertini2020finite}. We confirm this scaling in our numerical simulations (\figref{fig:schmidt}). Notice that the exponent $p$ increases slightly at lower temperatures, indicating a faster decay of the Schmidt coefficients, even though the proportionality factor is larger.  Therefore we expect the non-equilibrium state to have an efficient representation in term of tensor networks even at low temperatures. 

\begin{figure}[ht]
\begin{center}
\includegraphics[width=\columnwidth]{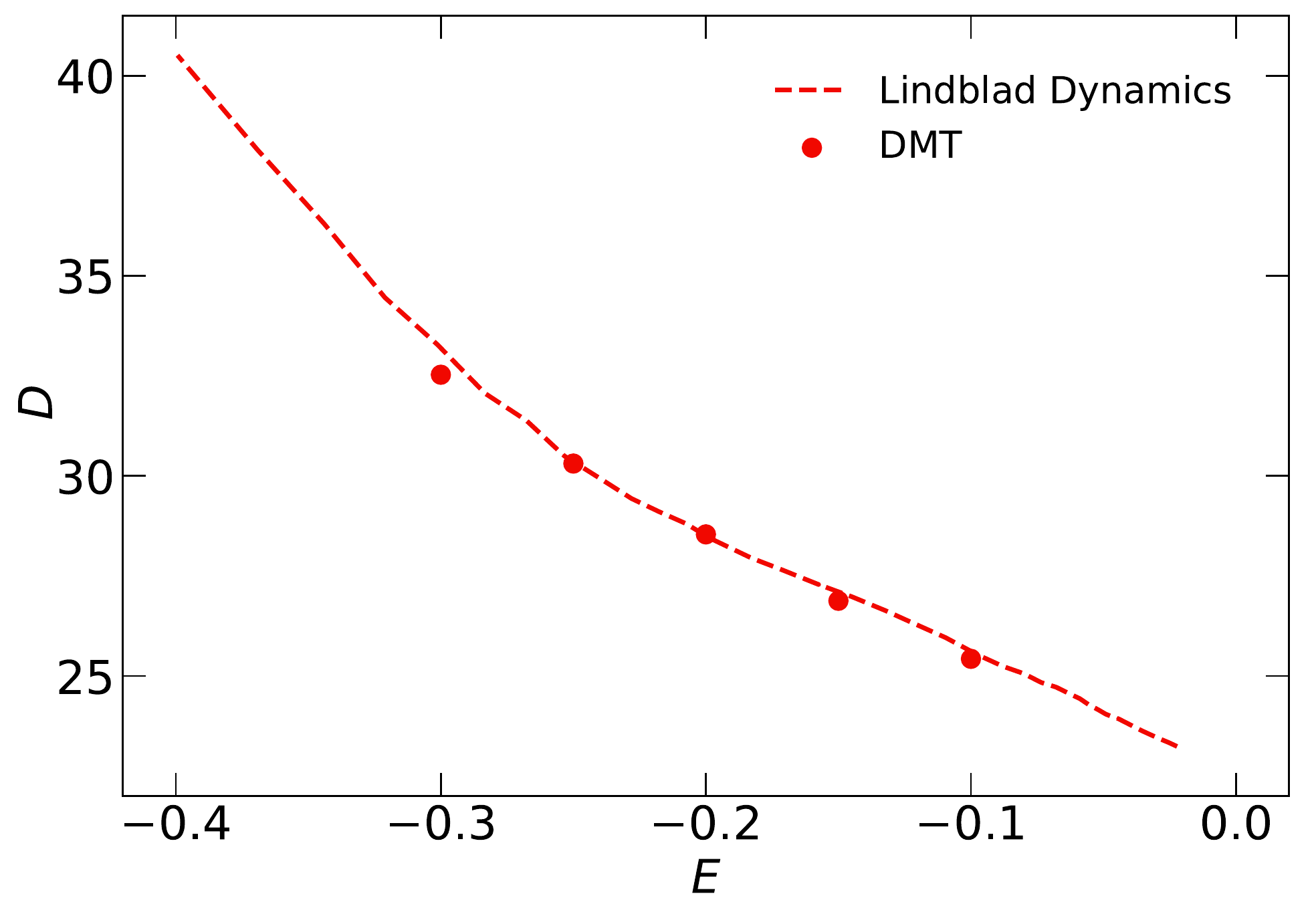}
\caption{Diffusion constant $D$ as a function of average energy density $E=\langle H_2\rangle/(L-1)$ estimated using Lindblad dynamics (dashed line) compared to the results from DMT calculations (dots, Ref.~\cite{ye2019emergent}). Both results are for an XZ chain of size $L=100$.}
\label{fig:DMT}
\end{center}
\end{figure}

In order to optimize our computational resources, we choose different $\chi$ at various stages in our evolution. We start with a large bond dimension $\chi=200$ during the early stages, when we have rapid entanglement growth. Then for intermediate times, the bond dimension is reduced to $\chi=64$, as the state continues to approach the NESS. Finally, at late times we increase the bond dimension back to $\chi=200$ in order to fine-tune our solution. We always check the convergence of NESS observables with the bond dimension and found $\chi=200$ to be sufficient. We also check that the NESS is independent of initial conditions by initializing our system with different states $\rho(0)$ and confirming that we always get the same final state $\rho_\infty$.


The expectation value of any observable in NESS can be computed as $\langle \mathcal{O} \rangle = \Tr(\mathcal{O}\rho_\infty)/\Tr(\rho_\infty)$. In particular, we are interested in on-bond energies $E_{i, i+1}=\langle H_{i, i+1} \rangle$ and currents $j_i$. The non-equilibrium state typically contains boundary effects due to the bath coupling. Hence we disregard the values of physical quantities close to the edges and only report our findings for the bulk values. Moreover, due to the inexact representation of $\rho_\infty$ as a tensor network, there are small fluctuations in the expectation values even at late times. Hence, we usually average our numerical results over $10^3$ Trotter steps $\delta t$. 

Certain tensor network methods, such as the time-dependent variational principle~\cite{Haegeman2011}, can converge quickly with bond dimension, but the resulting dynamics yield unphysical diffusion coefficients~\cite{leviatan2017quantum,kloss2018time}. We make sure that our results for the diffusion constants are correct by cross-checking them with other recent methods. For the Ising model in a tilted field, we compare our results to those obtained via dissipation-assisted operator evolution~\cite{rakovszky2020dissipation} and find perfect agreement, at least at high temperatures. Similarly, the diffusion constants reported for the XZ model agree with those derived using density matrix truncation (DMT)~\cite{white2018quantum,ye2019emergent} for a wide range of temperatures. A comparison of the two methods is shown in \figref{fig:DMT}, where we plot our data as a function of energy density to match the results presented in Ref.~\cite{ye2019emergent}.
\section{Local Thermal States}
\label{sec:appendixC}

We briefly review the approach to thermometry based on local thermal states, as described in Ref.~\cite{mendoza2015}, and comment on its relevance for our method in the main text. The key idea is to approximate the NESS reduced density matrix on two sites $\tilde{\rho}_{i, i+1}$ with a thermal density matrix $\rho_{i, i+1}$ by minimizing the trace distance $D(\tilde{\rho}_{i, i+1}, \rho_{i, i+1})$ (see \eqref{eq:trace_dist}). Now we have to determine a suitable ansatz for the density matrix $\rho_{i, i+1}$. We can start with a high-temperature expansion of a global thermal state~\cite{garcia2009,mendoza2015}, for which the two-site reduced density matrix is given by 

\begin{equation}
   \rho_{i, i+1} = \frac{e^{-H_{i, i+1}/T_{i, i+1}}}{\Tr\left(e^{-H_{i, i+1}/T_{i, i+1}}\right)}.
\end{equation}
However, this choice disregards the interactions between the two spins and the rest of the chain. Next, we can add a correction term that takes into account the neighboring spins in a mean-field fashion. A generic two-body interaction at the boundary can be approximated as 

\begin{equation}
    \sigma_{i-1}^\alpha\sigma_i^\beta \approx \langle\sigma_{i-1}^\alpha\rangle\langle\sigma_i^\beta\rangle + \sigma_{i-1}^\alpha\langle\sigma_i^\beta\rangle + \langle\sigma_{i-1}^\alpha\rangle\sigma_i^\beta,
\end{equation}
where $\alpha, \beta \in\{x, y, z\}$ and similarly for the other end $(i+1, i+2)$. If we replace the boundary operators by their mean field values, we end up with a site-dependent chemical potential term for each Pauli operator whose expectation value is non-zero $\langle\sigma^\alpha\rangle\neq 0$ and the ansatz state becomes 

\begin{equation}
   \rho_{i, i+1} = \frac{e^{-(H_{i, i+1}+\sum_{\alpha=x, y, z}(\mu_i^\alpha\sigma_i^\alpha+\mu_{i+1}^\alpha\sigma_{i+1}^\alpha))/T_{i, i+1}}}{\Tr\left(e^{-(H_{i, i+1}+\sum_{\alpha=x, y, z}(\mu_i^\alpha\sigma_i^\alpha+\mu_{i+1}^\alpha\sigma_{i+1}^\alpha))/T_{i, i+1}}\right)}.
   \label{eq:rho_thermal}
\end{equation} 
For the Ising model in a tilted field we have $\langle\sigma_i^y\rangle = 0$ and hence $\mu_i^y = 0$ for all $i$. Similarly, the symmetries of the XZ Hamiltonian impose $\langle\sigma_i^y\rangle =\langle\sigma_i^z\rangle = 0$  and therefore $\mu_i^y = \mu_i^z = 0$ for all $i$. To go beyond the mean-field approximation, the remaining chemical potentials are taken as fitting parameters, together with the temperature $T_{i, i+1}$. The optimization procedure is described in Ref.~\cite{mendoza2015}. Finally, we mention that the energy currents are not included in this two-site description because they involve three consecutive sites (see \eqref{eq:current_1} and \eqref{eq:current_2}).

A potential drawback of this method is its reliance on the thermal state ansatz in \eqref{eq:rho_thermal}, which is not always applicable. Even in the absence of a temperature gradient in the system, the predicted local temperature can be very different from the global temperature, especially in the low-temperature regime~\cite{garcia2009}. Moreover, the parameters of the thermal state are found either by solving the non-linear optimization problem directly or through an iterative self-consistent procedure~\cite{mendoza2015}. These methods can suffer from multiple local minima and can become inefficient if the parameter search space is large. 

Our method for extracting the local temperature described in \secref{sec:thermal} is designed to address these shortcomings. We expect our solution $\rho_2(T_{i, i+1})$ to capture the interactions with the rest of the system much better than a mean-field approximation and it has the
additional benefit of exactly recovering the global temperature of an equilibrium system. However, our solution is hard to interpret in terms of a local Hamiltonian, which on the other hand is possible with \eqref{eq:rho_thermal}. Therefore, the two methods can be used in tandem, and if they agree, we can gain additional insight into the local structure of the reduced density matrix solution. 

It is worth mentioning that both procedures are improvements over previous methods where the temperature was calculated by matching specific local observables (e.g. energy) with their expectation values in a thermal state~\cite{znidaric2010,vznidarivc2011transport}.

\section{Thermal Conductivity and Heat Capacity}
\label{sec:appendixD}

\begin{figure}[t]
\begin{center}
\includegraphics[width=\columnwidth]{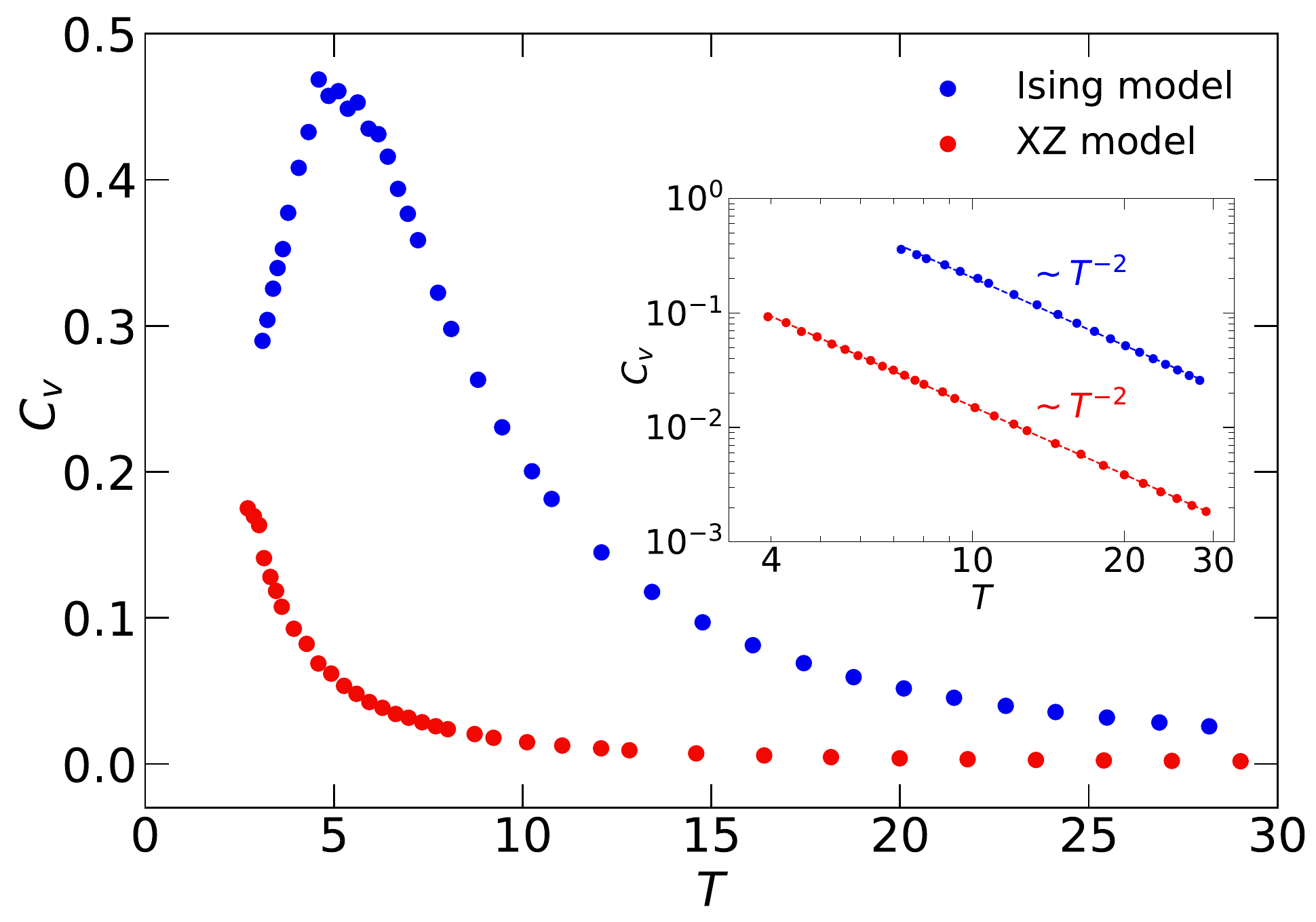}
\caption{Temperature dependence of the heat capacity $C_v$ for the tilted-field Ising model (blue) and the XZ model (red). The inset shows a quadratic fit to the heat capacity $C_v\sim T^{-2}$ at high temperatures.}
\label{fig:heat_capacity}
\end{center}
\end{figure}

In a system without charged excitations (which is the case for both our models), the energy and heat currents must be equal and hence we can define the thermal conductivity $\kappa$ through the relation $j=-\kappa\nabla T$. Having access to temperature gradients from our local temperature profiles, we are now in a position to extract the temperature dependence of the thermal conductivity. The connection between $\kappa$ and $D$ is captured by the Einstein relation $\kappa = C_vD$, where $C_v$ is the heat capacity. Depending on the experimental realization at hand, measuring the heat capacity or thermal conductivity may be preferred over the diffusion constant. Our methods provide access to all three of these quantities.  

It turns out that the diffusion constant and thermal conductivity have similar temperature dependencies, and it is more interesting to examine the heat capacity as their ratio. The heat capacity as a function of temperature for our two models is plotted in \figref{fig:heat_capacity}. At high temperatures, we expect $E\sim -1/T$ and therefore $C_v\sim T^{-2}$. The inset log-log plot clearly shows the heat capacity approaching zero quadratically at high temperatures. At low temperatures, on the other hand, the relevant energy scale is the excitation gap $\Delta$. At temperatures below $\Delta$, we expect the heat capacity to be exponentially small $C_v\sim e^{-\Delta/T}$, since the number of low-lying excitations is exponentially suppressed. This is visible for the tilted-field Ising model, for which we can reach temperatures $T\ll \Delta$. For the XZ model, we do not access low enough temperatures to witness a decrease in heat capacity. 

\end{document}